\documentclass[%
preprint,
superscriptaddress,
 amsmath,amssymb,
 aps,
]{revtex4-2}

\usepackage[utf8]{inputenc}
\usepackage{amsmath}
\usepackage{amsthm}
\usepackage{array}
\usepackage{graphicx}
\usepackage{natbib}
\usepackage{float}
\usepackage{color}
\usepackage{indentfirst}

\usepackage{xspace}
\usepackage{hyperref}

\newcommand{\spheading}[1]{%
\vspace{8pt}
\begin{centering}\MakeUppercase{\textbf{#1}}\\%
\end{centering}%
\vspace{8pt}%
}

\newcommand{\bpped}{\beta_{p,\mathrm{ped}}}        
\newcommand{\dped}{\Delta_{\mathrm{ped}}}          
\newcommand{\aped}{\alpha_{\mathrm{ped}}}          
\newcommand{\omstar}{\omega_*}                     
\newcommand{\gnorm}{\gamma/(\omstar/4)}            
\newcommand{\Wcmsq}{\mathrm{W/cm^2}}               

\begin{document}

\author{J. McClenaghan}
\thanks{Author to whom correspondence should be addressed. Email: mcclenaghanj@fusion.gat.com}
\affiliation{General Atomics, PO Box 85608, San Diego, CA 92186–5608, USA}

\author{K. E. Thome}
\affiliation{General Atomics, PO Box 85608, San Diego, CA 92186–5608, USA}
\author{T. F. Neiser}
\affiliation{General Atomics, PO Box 85608, San Diego, CA 92186–5608, USA}
\author{J. Candy}
\affiliation{General Atomics, PO Box 85608, San Diego, CA 92186–5608, USA}
\author{F. D. Halpern}
\affiliation{General Atomics, PO Box 85608, San Diego, CA 92186–5608, USA}
\author{T. H. Osborne}
\affiliation{General Atomics, PO Box 85608, San Diego, CA 92186–5608, USA}
\author{S. Saarelma}
\affiliation{UKAEA (United Kingdom Atomic Energy Authority), Culham Campus, Abingdon, Oxfordshire OX14
3DB, United Kingdom of Great Britain and Northern Ireland}
\author{the MAST-U team}
\affiliation{See Harrison et al 2019 (\url{https://doi.org/10.1088/1741-4326/ab121c}) for the MAST-U team}

\title{Integrating Gyrokinetic Flux Predictions with Ideal MHD Stability Boundaries}

\date{\today}

\begin{abstract}
Accurate prediction of pedestal height and width in tokamaks remains a critical issue as it strongly influences the predicted plasma performance of all future reactors. We present an integrated pedestal-stability workflow that combines equilibrium scans with magnetohydrodynamic (MHD) stability analysis using ELITE and GATO and gyrokinetic transport predictions using CGYRO/QLGYRO. The workflow reproduces the characteristic KBM first- and second-stability structure previously identified in gyrokinetic pedestal studies. Applied to spherical tokamaks (STs), the workflow shows good agreement with past studies when low-$n$ peeling stability is included, emphasizing the importance of resolving low-$n$ physics in ST pedestals. Extending the analysis to SPARC-like, high toroidal field plasmas, kinetic ballooning mode (KBM) and microtearing mode (MTM) heat fluxes are found to increase strongly with toroidal field, suggesting that access to the KBM second-stability region may become significantly more difficult in high toroidal field devices. Comparison with global ELITE finite-$n$ analysis at SPARC parameters suggests that the H-mode pedestal lies in an intermediate regime bounded by local KBM second stability on one side and global finite-$n$ ballooning instability on the other, consistent with the EPED picture once global effects are included.
\end{abstract}

\maketitle


\section{Introduction}\label{sec:intro}

Accurately predicting pedestal pressure is essential for reliable extrapolation of plasma performance to future fusion reactors. The pedestal height sets the core pressure through stiffness and profile coupling, and therefore strongly influences global confinement and fusion gain. A widely used predictive framework for H-mode pedestal structure is the EPED model \cite{snyder:2011}, which determines the pedestal height by combining peeling–ballooning stability limits (typically evaluated with ELITE \cite{snyder:2002, wilson:2002}) with a constraint on the normalized pedestal poloidal flux width derived from the kinetic ballooning mode (KBM) scaling, $\dped = \beta_{p,\mathrm{ped}}^{1/2} \, G$, where $\beta_{p,\mathrm{ped}}=p_{\mathrm{ped}}/(B_p^2/2\mu_0)$ is the local poloidal beta at the pedestal, $p_{\mathrm{ped}}$ is the thermal pressure at the pedestal, $B_p$ is the poloidal magnetic field, and $G$ is a weakly geometry- and collisionality-dependent factor. In the standard EPED picture, the KBM clamps the normalized pedestal pressure gradient: between ELMs the local gradient rises until the KBM is destabilized, after which further pedestal growth proceeds at approximately fixed normalized gradient with the height and width increasing together along the relation $\dped = \bpped^{1/2}\,G$. This expansion continues until the peeling--ballooning boundary is reached and an ELM is triggered, repeating the cycle.
For conventional, higher–aspect-ratio tokamaks, a value of $G \approx 0.076$ has provided good agreement with experiment.

However, there have been discrepancies in the height-width scaling on spherical tokamaks. In MAST-U \cite{harrison:2019}, experimentally inferred pedestals were significantly wider, corresponding to $G \approx 0.11$ \cite{knolker:2021}. Studies of NSTX \cite{ono:2000} spherical tokamak plasmas have reported even stronger deviations, with pedestal width scaling approximately linearly with $\bpped$ rather than following the square-root dependence \cite{diallo:2013}. In addition, for such wide pedestals, ELITE often predicts that the plasma lies well within typical stability thresholds, suggesting that either low toroidal mode number (low-$n$) limits or non-ideal effects may be playing a more significant role at low aspect ratio than assumed in the standard EPED paradigm.

Consistent with the EPED picture, pedestals in NSTX H-mode and enhanced pedestal H-modes were
found to be within 10\% of the local KBM threshold \cite{battaglia:2020}.
To address the KBM constraint more directly, the GKPED framework \cite{parisi:2024} employed gyrokinetic simulations to scan pedestal height and width and evaluate KBM stability self-consistently. One notable result from that work was that experimentally observed ELMy H-mode pedestals often appeared to reside in the second-stability region of the KBM. This raises an important question: if the pedestal is second-stable to the KBM, what mechanism is regulating the pedestal gradient and width?

In this work, we construct combined microturbulence-MHD stability maps to assess possible operating space for H-modes by scanning pedestal height and width using self-consistent equilibria and evaluating both gyrokinetic transport (via QLGYRO \cite{patel:2021}) and ideal MHD stability (via ELITE \cite{snyder:2002, wilson:2002} and GATO \cite{bernard:1981}). The QLGYRO fluxes are obtained from linear CGYRO simulations \cite{candy:2016}, combined with the TGLF SAT1 saturation rule \cite{staebler:2017}. The linear CGYRO runs are executed using the OMFIT GYRO\_GACODE module, which enables SLURM-based job management through OMFIT \cite{meneghini:2015}. This workflow allows the quasilinear flux calculations to be fully automated within OMFIT and is the approach used in this work.

These flux predictions are then overlaid with peeling–ballooning stability boundaries, enabling direct comparison between microinstability-driven transport and macroscopic MHD limits within the explored parameter space.

This combined approach allows us to reproduce the KBM stability structure observed in GKPED studies and identify narrow regions of KBM second stability that are constrained by high-$n$ peeling limits. By integrating microinstability transport predictions with MHD constraints, we provide a more complete picture of the pedestal operating space than can be obtained from either framework alone.

The remainder of this paper is organized as follows. Section~\ref{sec:ST} describes the construction of the stability maps and applies them to an NSTX-like $A=1.5$ plasma. How the stability maps change for slightly higher aspect ratios in NSTX-U and MAST-U is discussed in Section~\ref{sec:STU}. Section~\ref{sec:CON} examines the KBM stability map of a conventional tokamak, where we find that QLGYRO fluxes from KBMs become much stronger at high magnetic fields, discusses the potential implications for H-mode access, and uses ELITE finite-$n$ ballooning calculations to show how global effects close off the upper boundary of the KBM second-stable operating window. Section~\ref{sec:conclusion} summarizes the main conclusions.

\section{Pedestal Stability Map Construction for NSTX-Like Plasma}\label{sec:ST}

To create equilibria for the microturbulence-MHD stability maps, we use the OMFIT PRO\_create module \cite{slendebroek:2023} to generate a set of equilibria by scanning the pedestal height and width. The equilibria are constructed to be representative of NSTX-like plasmas like NSTX discharge 139047 \cite{diallo:2013}. Details of the global parameters of this plasma and other plasmas in the paper are given in Table~\ref{tab:devices} including aspect ratio ($A$), elongation ($\kappa$), magnetic field ($B_t$), plasma current ($I_p$), normalized beta $\beta_N=\beta a B_t/I_p$, and pedestal electron density $n_{e,\mathrm{ped}}$. The pedestal temperature is scanned over $T_{e,\mathrm{ped}} = 0.3–0.9$~keV, while the pedestal width is varied in the range $\dped = 0.075$–$0.3$. Additionally, all scans in this paper assume ion temperature $T_i=T_e$, separatrix temperature $T_{e,\mathrm{sep}}=70$~eV, and $Z_{\mathrm{eff}}=2$.

\begin{table}[h]
\centering
\caption{Parameters of the tokamak plasmas simulated in this work.}
\label{tab:devices}
\begin{tabular}{lcccccc}
\hline\hline
Device & $A$ & $B_t$ (T) & $I_p$ (MA) & $\kappa$ & $\beta_N$ & $n_{e,\mathrm{ped}}$ ($10^{20}$~m$^{-3}$) \\
\hline
NSTX-like           & 1.5  & 0.47 & 1.0 & 2.5   & 5 & 0.5 \\
NSTX-U              & 1.65 & 0.62  & 1.0  & 2.12  & 4 & 0.5 \\
NSTX-U (full shape) & 1.69 & 1.0  & 1.8 & 2.79 & 5 & 0.9 \\
MAST-U              & 1.56   & 0.59   & 0.75  & 2.07   & 2 & 0.3 \\
SPARC (2~T)         & 3.2  & 2.0  & 1.4  & 1.9   & 1 & 0.52 \\
SPARC (12~T)        & 3.2  & 12.2 & 8.75  & 1.9   & 1 & 3.3 \\
\hline\hline
\end{tabular}
\end{table}

The toroidal current profile is prescribed as

\begin{equation}
J = J_0 (1 + \psi_n^\alpha)^\beta + J_{\mathrm{bs}} ,
\end{equation}

\noindent where $J$ is the flux surface averaged toroidal current density $\langle J_t/R \rangle/\langle 1/R\rangle$, $J_t$ is the toroidal current density, $R$ is the major radius, $\psi_n$ is the normalized poloidal flux, $J_{\mathrm{bs}}$ is the bootstrap current calculated using the Redl model \cite{redl:2021}, and $J_0$ is adjusted to match the specified total plasma current. The shaping parameters $\alpha$ and $\beta$ are unity by default and are varied as needed to ensure that the on-axis safety factor satisfies $q_0 > 1$.

\begin{figure}
    \centering
    \includegraphics[width=3.5in]{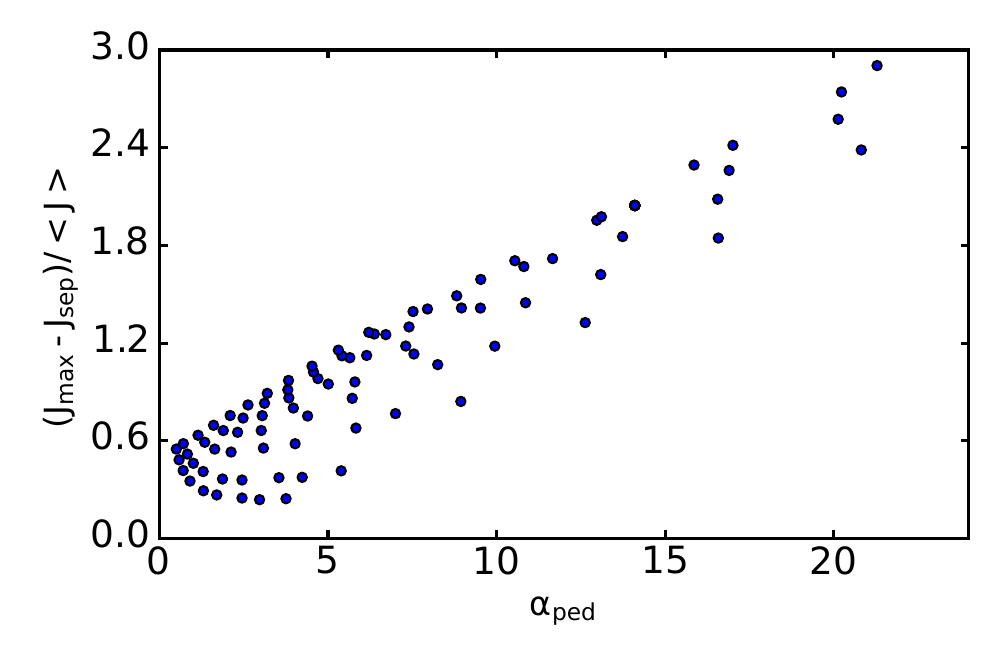} \\
    \includegraphics[width=3.5in]{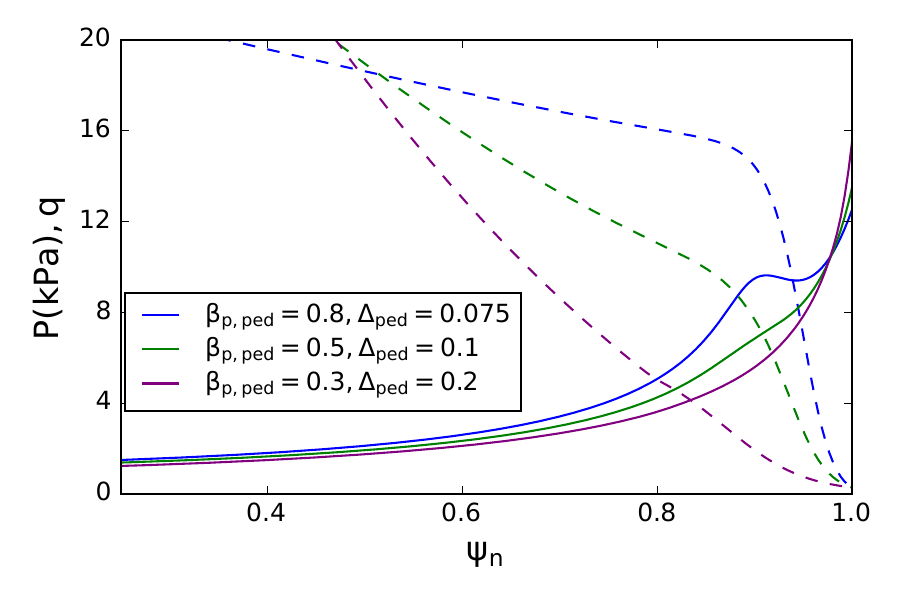} 
    \caption{Scanned pedestal values are plotted on typical VARYPED grid of $\aped$ and $(J_{\mathrm{ped}}-J_{\mathrm{sep}})/ J_{\mathrm{avg}}$ (top). The equilibrium $p$ (dashed) and $q$ (solid) profiles plotted vs. $\psi_n$ for select points (bottom).}
    \label{fig:to_varyped}
\end{figure}

Because the total plasma current is constrained to align with the bootstrap current which scales with the pressure gradient, the present scan differs from conventional peeling-ballooning scans performed with ELITE, where the pressure gradient and current density are varied independently.
To show this distinction, Figure~\ref{fig:to_varyped} shows the scan projected onto the standard peeling--ballooning parameter space scanned by VARYPED \cite{osborne:2015}, defined by the normalized pedestal pressure gradient $\aped = -\frac{2 \partial V/\partial \psi}{(2\pi)^2} \left(\frac{V}{2\pi^2 R}\right)^{1/2} \mu_0 \partial p/ \partial\psi$ \cite{miller:1998}, where $p$ is the plasma pressure, $\psi$ is the poloidal flux, $V$ is the plasma volume, and $R$ is the major radius, and the normalized edge current density $(J_{\mathrm{ped}} - J_{\mathrm{sep}})/ J_{\mathrm{avg}}$, where $J_{\mathrm{ped}}$ is the current density at the pedestal, $J_{\mathrm{sep}}$ is the current density at the separatrix, and $J_{\mathrm{avg}}$ is the average current density across all flux surfaces. Although some spread is present, most equilibria fall approximately along the relation:

\begin{equation}
\aped \approx 0.15\, \frac{J_{\mathrm{ped}} - J_{\mathrm{sep}}}{J_{\mathrm{avg}}}.
\end{equation}

The bottom panel of Figure~\ref{fig:to_varyped} compares pressure and safety factor profiles for three pedestal configurations with different pedestal heights $\bpped$ and widths $\dped$. Increasing pedestal height and reducing pedestal width produce a steeper edge pressure gradient and stronger modification of the edge current profile. Broader, lower-pressure pedestals exhibit a monotonic $q$-profile. At very high $\beta_{p,\mathrm{ped}}=0.8$ and for narrow pedestals of $\dped=0.075$, the bootstrap current substantially reduces the edge magnetic shear, leading to flattening or even slight reversal of the $q$-profile. As we will show, large edge current densities and low magnetic shear ($s=\frac{r}{q}\frac{dq}{dr}$, where $r$ is the Miller minor radius) are unstable to peeling modes. 

\begin{figure}
    \centering
    \includegraphics[width=3.5in]{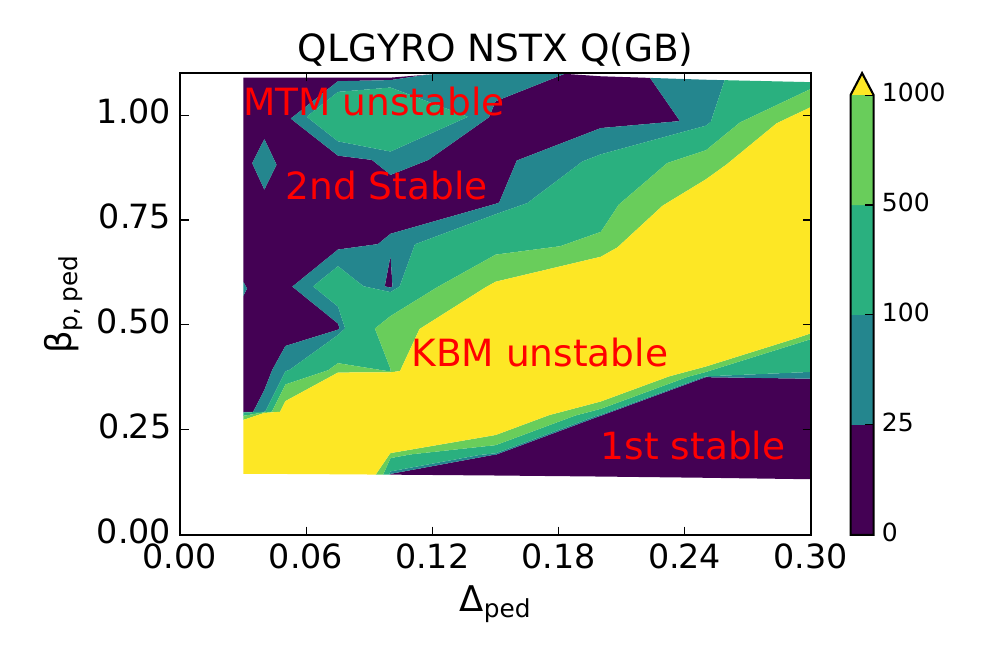} \\
    \includegraphics[width=3.5in]{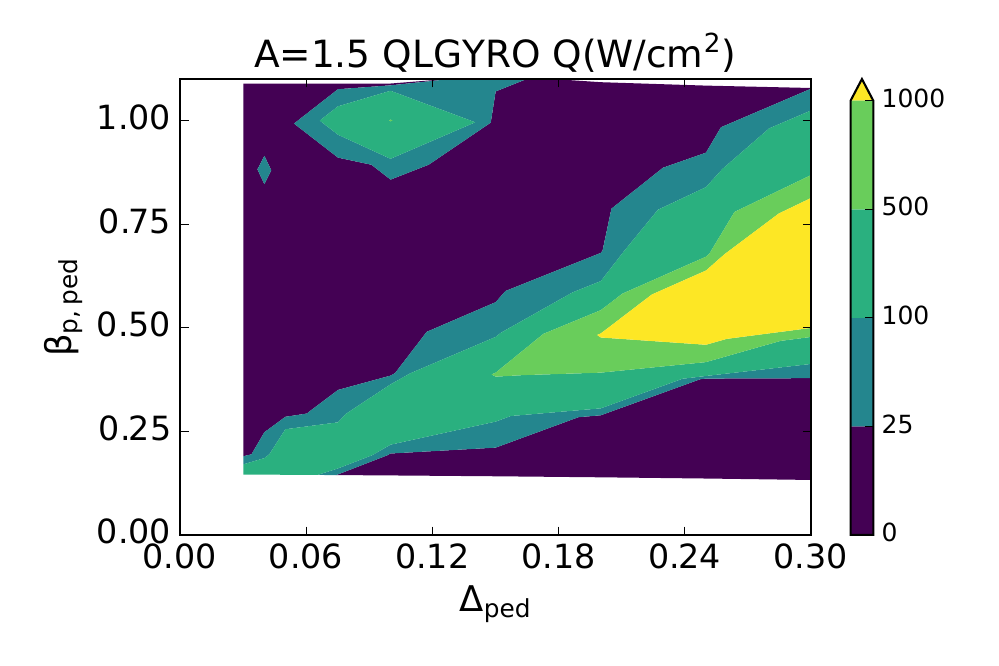} 
    \caption{QLGYRO predicted fluxes in $A=1.5$ NSTX-like plasma scanning $\bpped$ and $\dped$ in Gyro-Bohm units (top), and in physical units (bottom).}
    \label{fig:QLGYRO}
\end{figure}

We next examine QLGYRO stability while scanning the width and height of the pedestal. For consistency, simulations are performed at the same wavenumbers as a typical GKPED simulation of $k_\theta \rho_s = k_y = [0.06,\,0.12,\,0.18]$, where $k_\theta$ is the poloidal wavenumber and $\rho_s$ is the ion sound radius, and using the same radial grid (five nodes that extend from $0.25\,\dped$ to $0.75\,\dped$) as in a typical GKPED workflow. Linear CGYRO simulations are computed at each grid point, after which the TGLF SAT1 saturation rule is applied to obtain quasilinear fluxes.
Figure~\ref{fig:QLGYRO} shows the average electron energy flux across the five radial zones. A broad band of large heat flux ($Q>1000$~GB, where GB denotes CGYRO GyroBohm (GB) units energy $Q_{\mathrm{GB}} = n_e T_e c_s \rho_{*}^2$, $\rho_{*} = \rho_{s}/a$, $\rho_{s} = c_{s}/\Omega_{\mathrm{unit}}$, $c_s=\sqrt{e T_e/m_i}$, $\Omega_{\mathrm{unit}} = \frac{e B_{\mathrm{unit}}}{m_i c}$, $B_{\mathrm{unit}}(r) = \frac{q}{r} \frac{d\psi}{dr}$ , $a$ is the minor radius of the last closed flux surface, and $m_i$ is the deuterium mass) is observed, driven by KBMs due to their large growth rates. This structure is consistent with previous GKPED scans, where KBMs are the dominant instability along the diagonal. The scan separates naturally into four regions moving from lower right to top left: a first-stable KBM region, a KBM-unstable band, a second-stable KBM region, and an MTM-unstable region at high pedestal drive.  All scans in this paper will have these distinct regions, and vary only in details. 
Additionally, while unstable microtearing modes (MTMs) appear in the upper-left corner of the scan, these high-flux MTM cases lie above the ideal MHD stability boundary, making their physical relevance in this specific case unclear.

While plotting the fluxes in gyroBohm units facilitates identification of the KBM and MTM stability bands, conversion to physical units ($\Wcmsq$), shown in the bottom panel of Figure~\ref{fig:QLGYRO}, is more informative. In physical units, the predicted KBM-driven energy flux decreases significantly along the KBM band as the pedestal height and width are reduced.
At $\bpped = 0.6$ and $\dped = 0.3$, the predicted electron energy flux exceeds $500~\Wcmsq$, substantially above typical experimental values. In contrast, for a weaker pedestal with $\bpped = 0.25$ and $\dped = 0.08$, the predicted KBM-band energy flux is reduced to $170~\Wcmsq$, much weaker but still much larger than experimental levels.

The absolute values of the predicted heat fluxes should not be interpreted quantitatively, as the fluxes are primarily calibrated to ITG turbulence levels and KBM turbulence in local flux-tube simulations often does not saturate at realistic amplitudes. Nevertheless, the trends provide a meaningful qualitative indicator of KBM activity at given gradients.
In particular, the strong increase in $Q$ within the KBM-unstable region reflects the rapid growth of the underlying modes and underscores the relative difficulty of accessing the second stability region from the first. This indicates where the pedestal is likely to encounter MTM or KBM constraints and how sharply these limits restrict the achievable pedestal width and pressure.

\begin{figure}
    \centering
    \includegraphics[width=3.5in]{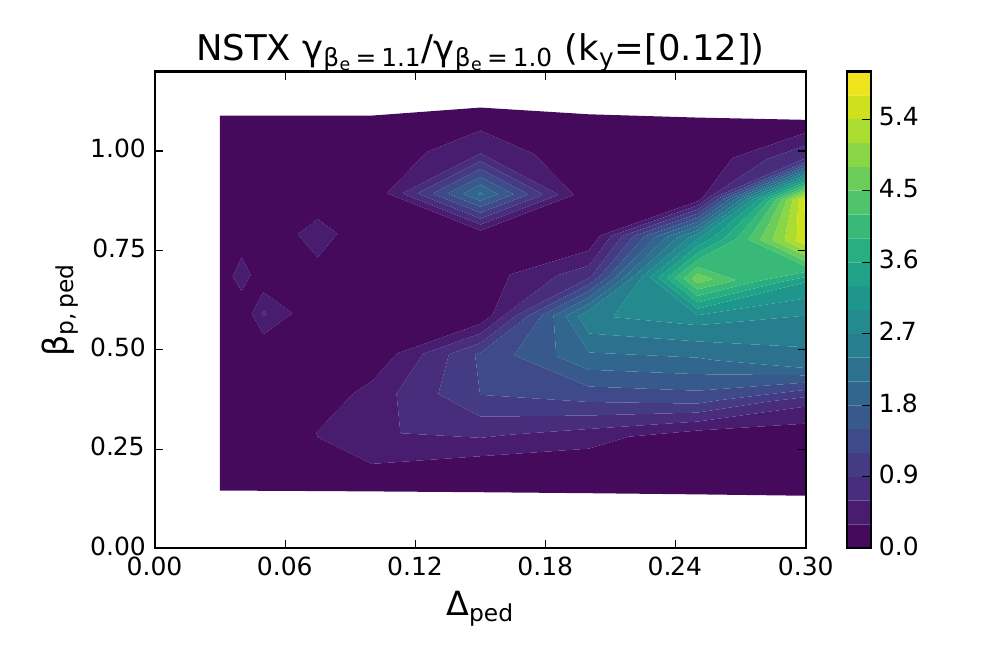} 
    \caption{Ratio of CGYRO growth rates $\gamma(\beta_e = 1.1\,\beta_{e,0}) / \gamma(\beta_e = \beta_{e,0})$ for the $k_\theta\rho_s = 0.12$ mode scanning in $\bpped$ and $\dped$ for the $A=1.5$ NSTX-like plasma.}
    \label{fig:betae_scan}
\end{figure}

As KBM growth rates are very sensitive to the electron beta $\beta_e=n_e T_e/(B_{\mathrm{unit}}^2/2\mu_0)$, we perform a typical $\beta_e$ sensitivity test across the scan to verify that the high-flux modes are indeed KBMs (or possibly hybrid trapped electron modes-KBMs). Figure~\ref{fig:betae_scan} shows the ratio of growth rates computed with $\beta_e$ artificially increased by 10\% to the growth rate at nominal value $\beta_{e,0}$, where $\beta_{e,0}$ is the value obtained from the equilibrium profiles.
Since KBM growth rates typically increase with $\beta_e$, a strong enhancement in the growth rate is indicative of a KBM. The regions that have a large heat flux in Figure~\ref{fig:QLGYRO} coincide with the same $\bpped, \dped$ locations where the growth rates increase significantly with $\beta_e$, providing further evidence that these modes are KBMs.

\begin{figure}
    \centering
    \includegraphics[width=3.5in]{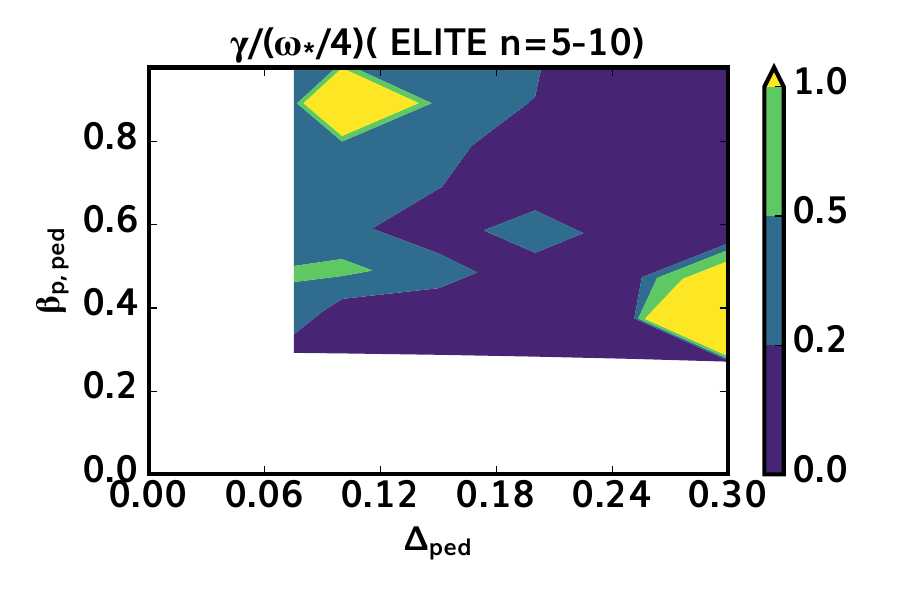} \\
    \caption{Maximum growth rate $\gnorm$ is plotted versus $\bpped$ and $\dped$ for ELITE $n=5$--$10$ modes.}
    \label{fig:mhd_elite}
\end{figure}

Since ideal global MHD stability sets the pedestal limit before an ELM, we next evaluate the MHD stability of the scanned equilibria. Figure~\ref{fig:mhd_elite} shows the predicted stability for toroidal mode numbers $n = 5$--$10$. Here we use a stability threshold of $\gnorm = 1$, where $\gnorm$ is the growth rate normalized to the maximum pedestal ion diamagnetic drift frequency ($\omega_{*i}=\frac{k_y T_i}{e B}\frac{d \ln p_i}{dr}$, where \(\omega_{*i}\) is the ion diamagnetic frequency, \(k_y\) is the poloidal wavenumber, \(T_i\) is the ion temperature, \(B\) is the magnetic field strength, \(p_i\) is the ion pressure, and \(r\) is the minor radial coordinate.) to account for effective stabilization from drift-wave transport.

ELITE predicts unstable peeling–ballooning modes primarily in the upper-left region of the scan above the line intersecting $\bpped = 0.3$, $\dped = 0.06$ and $\bpped = 1.0$, $\dped = 0.2$.
In addition, ELITE predicts unstable modes in the lower-right corner of the parameter space at $\bpped = 0.08$ and $\dped = 0.3$.

\begin{figure}
    \centering
    \includegraphics[width=3.5in]{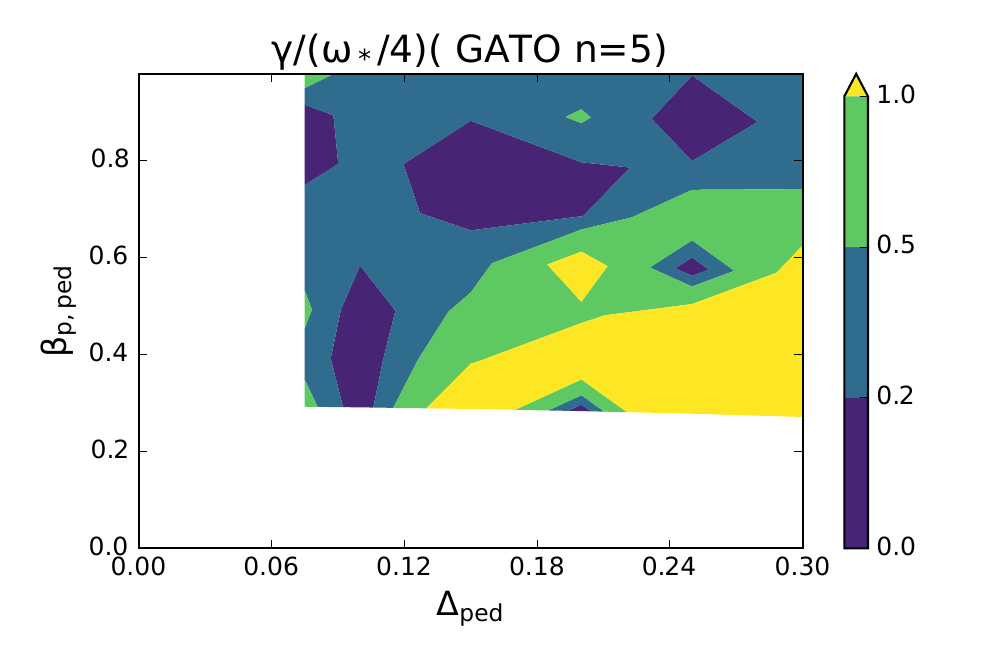} \\ 
    \includegraphics[width=3.5in]{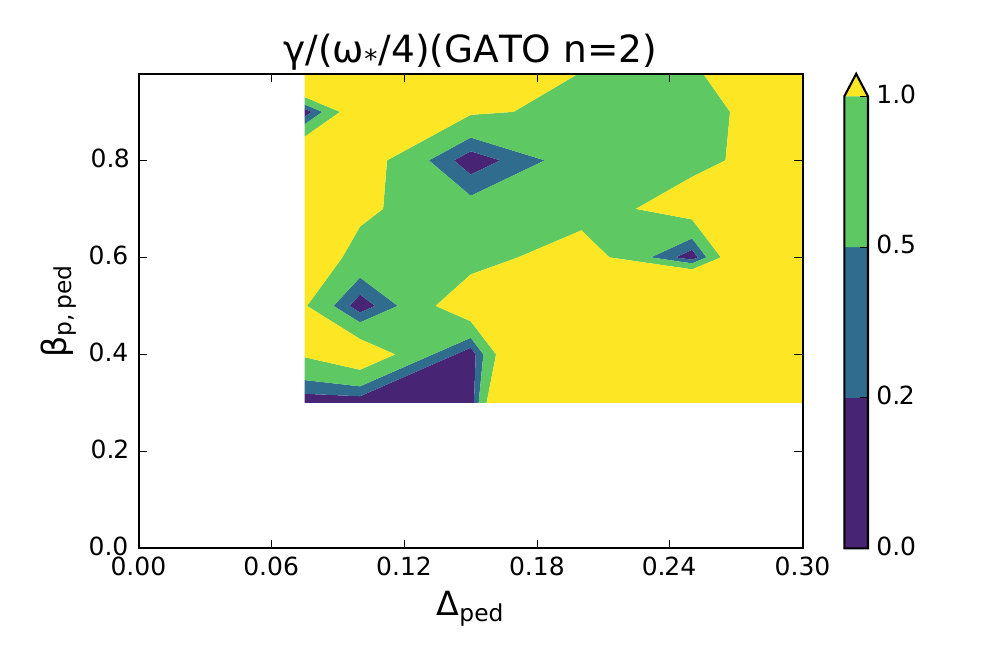}
    \caption{GATO maximum growth rates $\gnorm$ without filtering internal modes, plotted versus $\bpped$ and $\dped$ for $n=5$ (top) and $n=2$ (bottom).}
    \label{fig:mhd_gato_nofilter}
\end{figure}

We next examine the ideal MHD stability using GATO, which can better capture the stability of low toroidal mode number MHD modes ($n<5$). To be consistent with the ELITE calculations, no wall stabilization effects are included in GATO. In contrast to ELITE, GATO predicts increased instability at low $\bpped$ and large $\dped$.  The top panel of Figure~\ref{fig:mhd_gato_nofilter} shows the stability of the $n=5$ mode (i.e. the lowest ELITE predicted mode). The most unstable equilibria are located in the lower-right corner of the scan, with the only cases satisfying $\gnorm > 1$ occurring in this region. Shown in the bottom panel of Figure~\ref{fig:mhd_gato_nofilter}, for a lower toroidal mode number of $n=2$, GATO predicts broader instability across parameter space, with most equilibria unstable except those in the lower-left corner $\bpped=0.3$ and $\dped=0.06$--$0.12$. However, a diagonal band of reduced instability is observed, where the growth rates satisfy $\gnorm \approx 0.5$--$1$.

\begin{figure}
    \centering
    \includegraphics[width=3.5in]{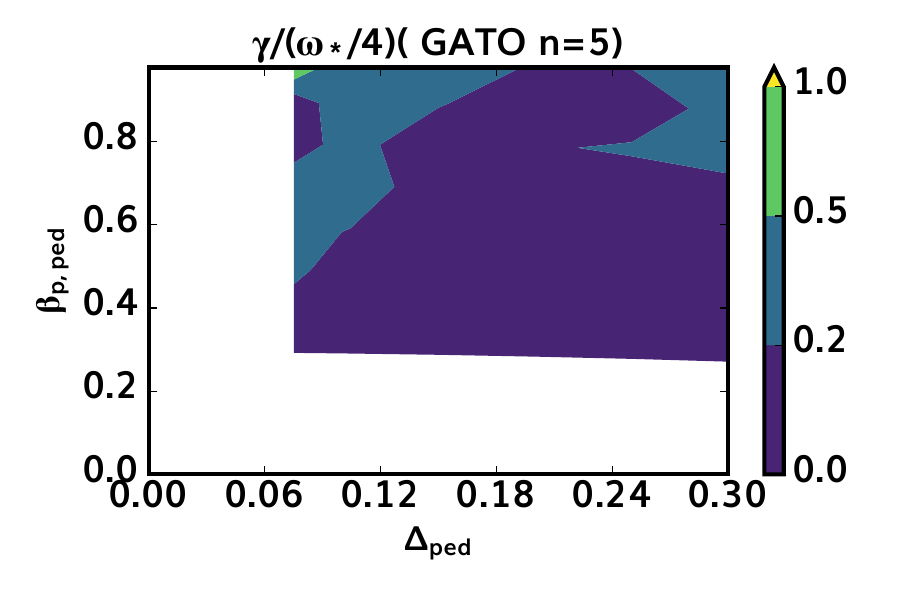} \\ 
    \includegraphics[width=3.5in]{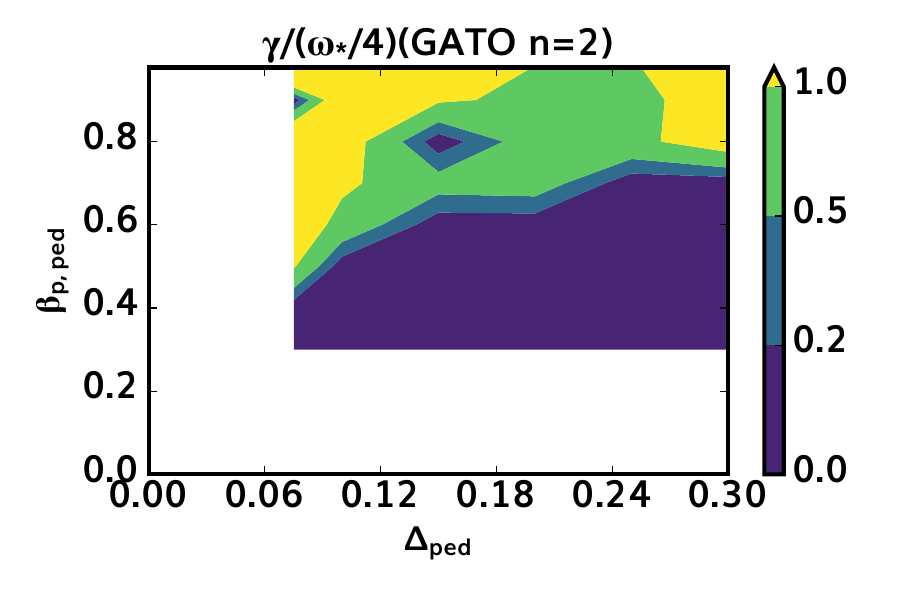}
    \caption{GATO maximum growth rate $\gnorm$ with internal modes filtered out, plotted versus $\bpped$ and $\dped$ for $n=5$ (top) and $n=2$ (bottom).}
    \label{fig:mhd_gato}
\end{figure}

After filtering out internal modes, the GATO results more closely resemble the ELITE predictions. To isolate peeling modes, eigenfunctions with peak flux surface average of the absolute value of the displacement $\xi$ located inside $\psi_n = 0.6$ are filtered out,  which corresponds to just outside the $q=2$ surface and removes core-localized modes and isolates edge-peeling structures.
The top panel of Figure~\ref{fig:mhd_gato} shows the $n=5$ peeling modes computed with GATO filtering out internal modes, which have a stability boundary similar to that obtained with ELITE. In particular, unstable modes are found in the upper left-hand corner near $\dped = 0.1$ and $\bpped = 0.8$, with growth rates satisfying $\gnorm > 0.2$.

For the lower-$n$ case, the modes are significantly more unstable for pedestals that occur at wider and higher $\bpped$. At $\dped = 0.1$ and $\bpped = 0.6$, the growth rate is approximately $ \gnorm =0.5$ and remains roughly constant as the pedestal width increases. Overall, ELITE and GATO predictions agree well at moderate toroidal mode numbers $n=5$, and ELITE is likely sufficient for sharp pedestals. However, for very wide pedestals, which can sometimes occur in NSTX, lower-$n$ toroidal modes should be included.

\begin{figure}
    \centering
    \includegraphics[width=3.5in]{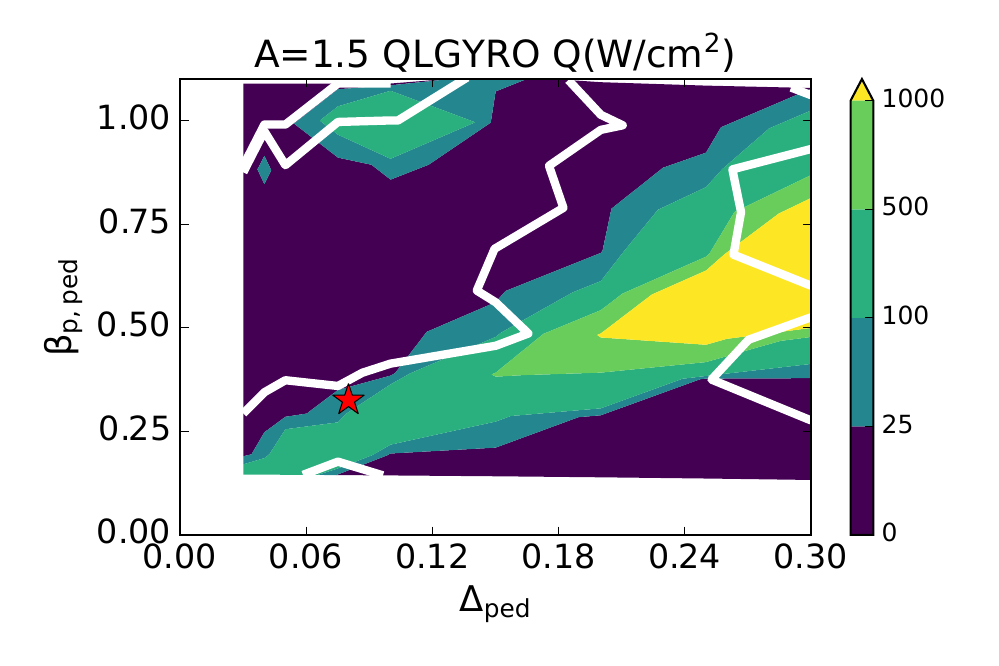} \\
    \caption{QLGYRO predicted fluxes combined with MHD threshold $\gnorm = 0.2$ in $A=1.5$ NSTX-like plasma scanning $\bpped$ and $\dped$.}
    \label{fig:qlgyro_wstability}
\end{figure}

Combining the QLGYRO flux predictions with the ideal MHD stability results yields a composite picture of the accessible H-mode operating space. Figure~\ref{fig:qlgyro_wstability} shows the QLGYRO heat fluxes with the MHD stability boundary overlaid in white. A narrow region emerges in which the pedestal is both second-stable to the KBM and stable to ideal MHD modes. This is consistent with a similar NSTX discharge 139047 from Diallo et al.~\cite{diallo:2013} which had a pedestal $\bpped = 0.25, \dped=0.08$. 
However, this result raises questions regarding consistency with the conventional EPED paradigm, in which the KBM constrains the ratio of pedestal height to width. If the local KBM is second-stable, the mechanism constraining the pedestal gradient remains unclear. One possibility is that higher $k_y$ modes not simulated here, like electron temperature gradient (ETG) modes are limiting the gradient. Another possibility is that the KBM is limiting the gradient, but global effects are needed for the correct constraint. This will be explored in more detail in section~\ref{sec:CON}.

\section{Spherical tokamaks MAST-U and NSTX-U}\label{sec:STU}

\begin{figure}
    \centering
    \includegraphics[width=3.5in]{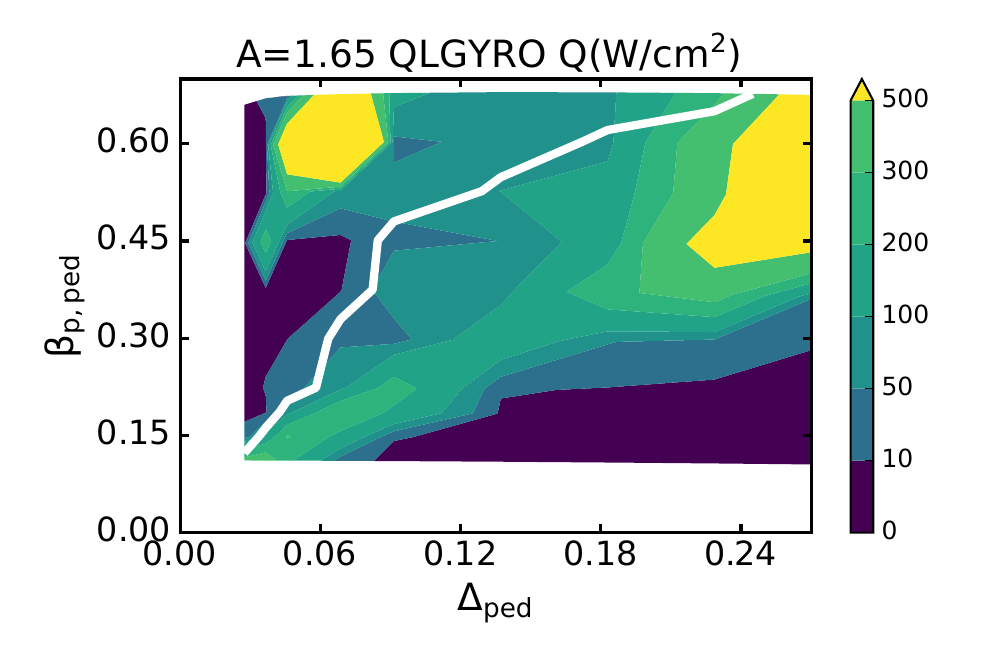} \\
    \includegraphics[width=3.5in]{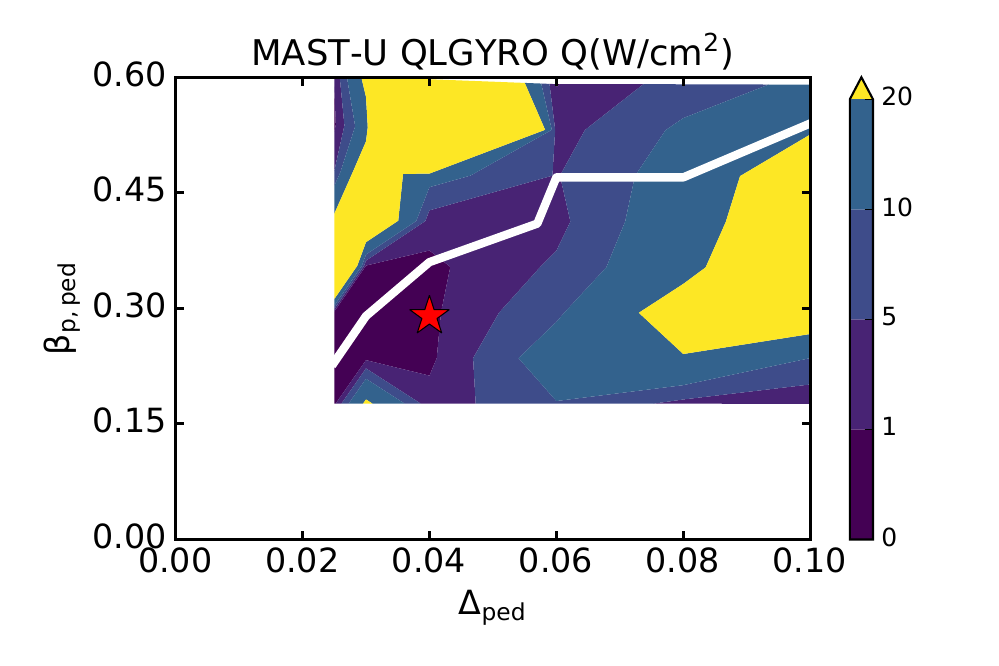}
    \caption{QLGYRO predicted fluxes combined with MHD threshold ($\gnorm = 0.2$) scanning $\bpped$ and $\dped$ for $A=1.65$ NSTX-U-like plasma (top) and MAST-U plasma (bottom).}
    \label{fig:nstxu_mastu}
\end{figure}

Up to this point, we have considered only the $A = 1.5$ case. We now examine slightly higher aspect ratio configurations. The first case corresponds to an NSTX-U–like plasma (discharge 204112) with $A = 1.65$, $\beta_N = 4$, and elongation $\kappa = 2.2$.
The top panel of Figure~\ref{fig:nstxu_mastu} shows the QLGYRO heat fluxes with the ideal MHD stability boundaries from ELITE and GATO overlaid. As in the $A = 1.5$ case, a narrow region emerges in which the pedestal is both second-stable to the KBM and stable to ideal MHD modes.
The bottom panel of Figure~\ref{fig:nstxu_mastu} shows the prediction for MAST-U discharge 45272, whose pedestal was analyzed in Imada \textit{et al.}~\cite{imada:2024}. As the experimental pedestal width was on the order of $\dped = 0.05$, the present scan explores only $\dped = 0.025$--$0.1$. The experimental operating point falls within the second-stable region, consistent with the analysis performed by Imada et al. However, the peeling boundary of $\bpped=0.35$ at $\dped=0.04$ is somewhat above the experimental pedestal height, which was $\bpped=0.28$. One potential explanation is that non-ideal MHD effects, which have been found to be important in spherical tokamaks \cite{kleiner:2025, liu:2025, pankin:2025}, are neglected in the present analysis. 

Next, we look at a projected high-performance NSTX-U plasma with the full NSTX shape with $B_t=1$~T, $I_p=1.8$~MA, $\kappa=2.79$, $A=1.69$ based on TRANSP runid:1123K55. The point at which the ELITE stability limit intersects the KBM boundary is at $\bpped=0.23$ and $\dped=0.07$. With the higher $B_t$ and $I_p$, this corresponds to a fourfold increase in pedestal pressure.

\begin{figure}
    \centering
    \includegraphics[width=3.5in]{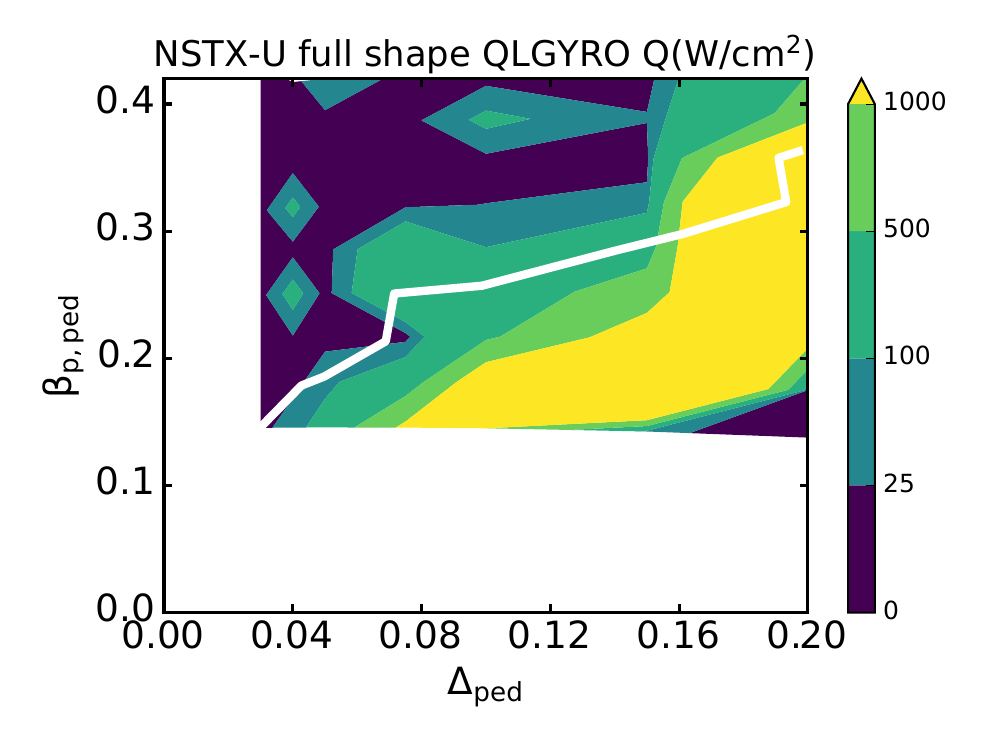}
    \caption{QLGYRO predicted fluxes combined with MHD threshold ($\gnorm = 0.2$) scanning $\bpped$ and $\dped$ for the $A=1.65$ NSTX-U-like plasma with full shape.}
    \label{fig:nstxu_fullshape}
\end{figure}

\section{Conventional Tokamaks}\label{sec:CON}

In this section, we examine how the stability map compares with that of a more conventional tokamak with a higher aspect ratio. To do so, we simulate a plasma at the ``primary reference discharge" SPARC parameters \cite{rodriguez:2022} with $A=3.2$ scanning the magnetic field from current device conditions of $B_t=2$~T like DIII-D \cite{luxon:2002} or ASDEX-U \cite{keilhacker:1985} to full field of $12.2$~T. We find that the general picture of the stability map is qualitatively the same as STs.

\begin{figure}
    \centering
    \includegraphics[width=3.5in]{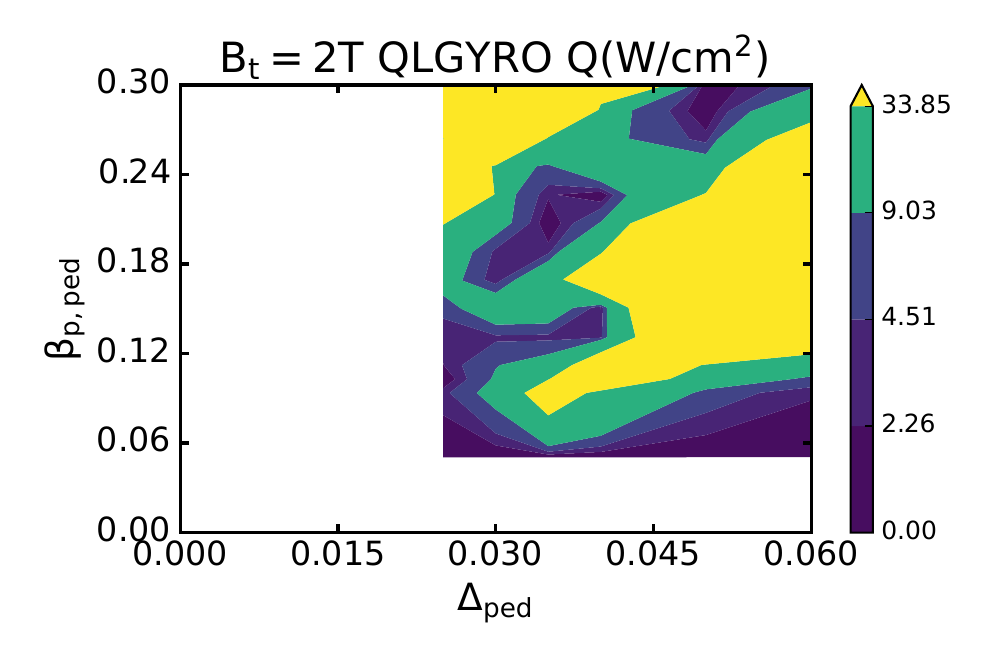}\\
    \includegraphics[width=3.5in]{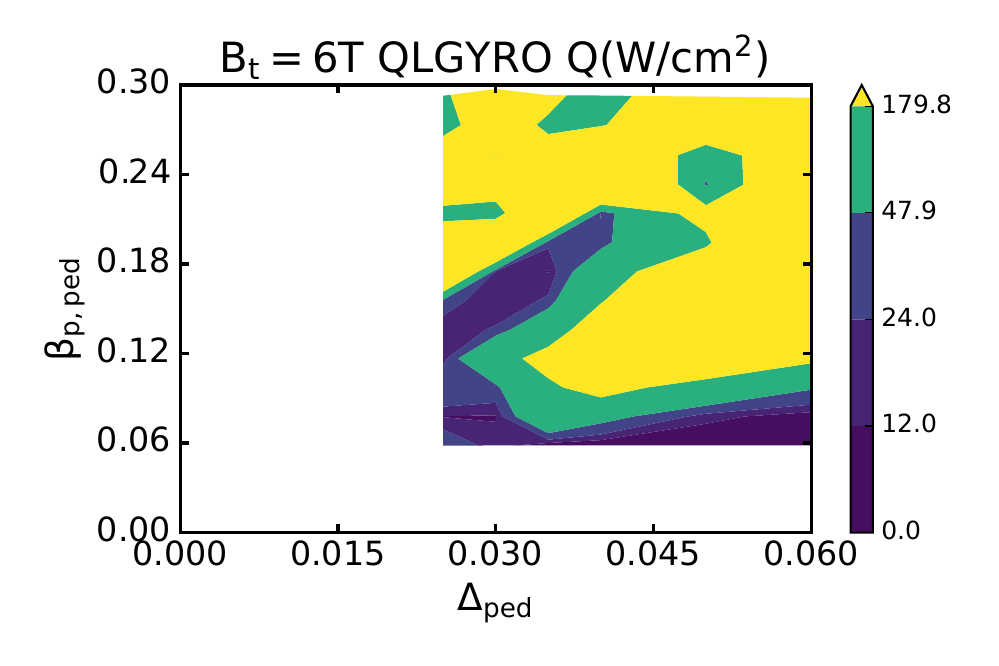} \\
    \includegraphics[width=3.5in]{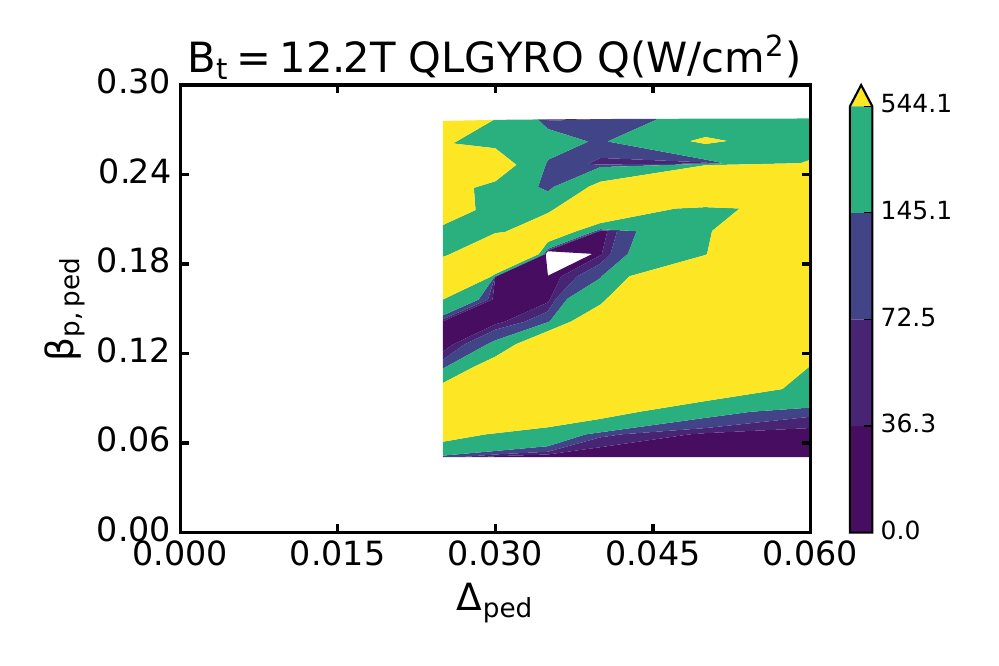} \\
    \caption{QLGYRO predicted fluxes scanning $\bpped$ and $\dped$ for SPARC-like 2~T plasma (top), 6~T plasma (middle), and 12~T plasma (bottom).}
    \label{fig:sparc}
\end{figure}

While the picture does not change significantly with $B_t$, the predicted fluxes from KBMs increase substantially. Figure~\ref{fig:sparc} shows QLGYRO stability scans for three different toroidal magnetic fields. As $B_t$ increases, $I_p$ and $n_{e,\mathrm{ped}}$ are linearly scaled to preserve the edge safety factor and to maintain a constant Greenwald fraction $n_{e,\mathrm{ped}}/n_{\mathrm{GW}}$, where $n_{\mathrm{GW}}=I_p/\pi a^2$. We note that the choice to scale by the empirical Greenwald density, as is typical in reactor projections, leads to lower collisionality at the higher magnetic fields for a given $\bpped$.
All three cases exhibit qualitatively similar stability structure to that observed in the spherical tokamak configurations: a first-stability region in the lower-right corner, followed by a KBM-unstable band, a KBM second-stable region, and finally an MTM-unstable region at higher pedestal drive.
Noting that the contour scales differ between panels, one key difference among the stability maps is that the QLGYRO-predicted KBM fluxes increase substantially with increasing $B_t$ (with $n_e$ and $I_p$ scaled accordingly).  
At $B_t = 2$~T, QLGYRO predicts KBM-driven fluxes of approximately $Q \approx 9~\Wcmsq$. 
Increasing the magnetic field to $B_t = 6$~T, the predicted flux rises to $Q \approx 50~\Wcmsq$. At the full SPARC field of $B_t = 12$~T, the KBM-driven flux further increases to  $Q > 500~\Wcmsq$.

\begin{figure}
    \centering
    \includegraphics[width=3.5in]{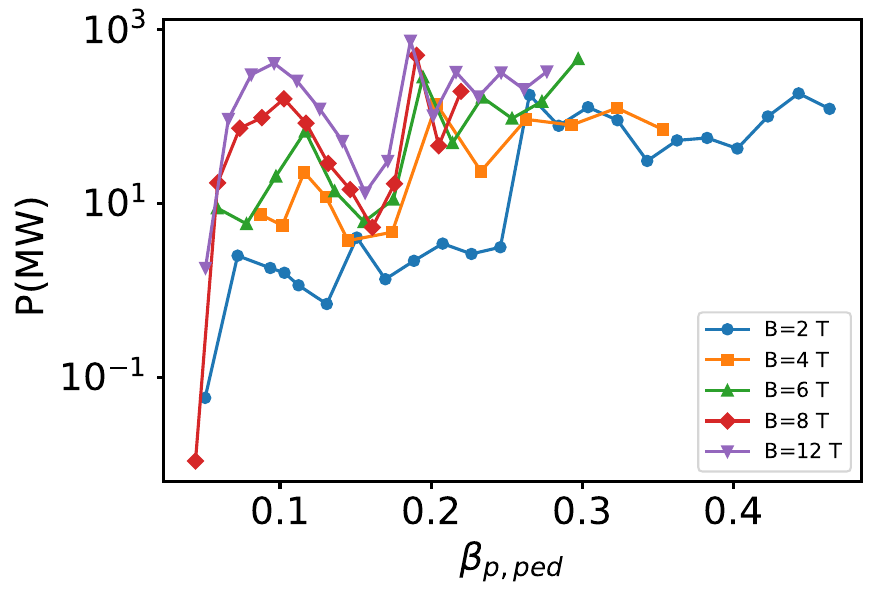}\\
    \includegraphics[width=3.5in]{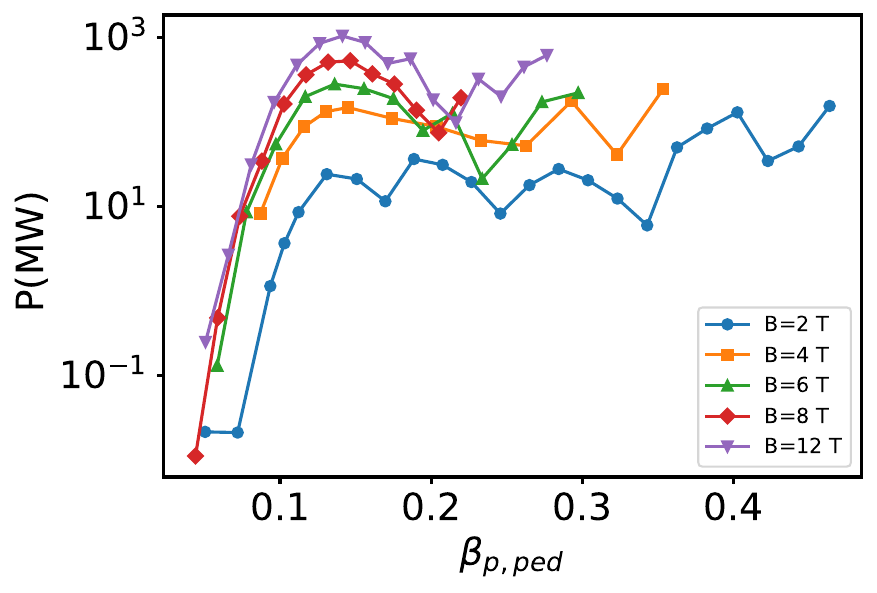}\\
    \includegraphics[width=3.5in]{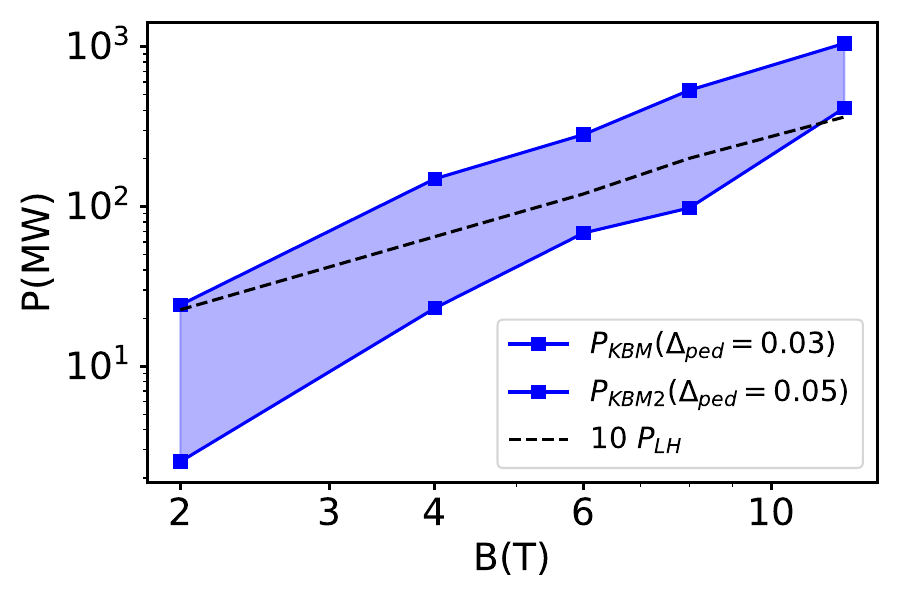}
    \caption{ QLGYRO predicted heat flux versus $\bpped$ at fixed $\dped = 0.03$ (top) and $\dped = 0.05$ (middle) for the SPARC-like scanning magnetic field. (Bottom) Peak power required to cross the KBM barrier versus $B_t$.}
    \label{fig:plh}
\end{figure}

To more clearly show the predicted power required to move from first stability to second stability of the local KBM barrier, the top panel of Figure~\ref{fig:plh} shows a slice at fixed width with the simple assumption of $\dped = 0.03$. All three magnetic field cases exhibit distinct first and second stability regions. The peak power required to cross the KBM barrier is approximately 10~MW at $B_t = 2$~T, 30~MW at $B_t = 6$~T, and 900~MW at $B_t = 12$~T.

Since QLGYRO predicts a finite energy flux threshold to transition from the first-stability (L-mode–like) region to the second-stability (H-mode–like) region, and this threshold increases with $B_t$ (and the correspondingly scaled $I_p$ and $n_e$), it is natural to ask whether the power required to access H-mode could be interpreted as the power needed to push through the KBM barrier.
As discussed earlier, the predicted heat-flux values should not be taken quantitatively. Nevertheless, the trends in the predicted fluxes may still provide insight into the relative accessibility of the second-stable regime. For example, the predicted power may be related to the power needed to transiently push through the before the KBM can grow in amplitude and flatten the pressure profile. 

\noindent The ITPA scaling for the L--H transition power threshold from Martin et al. \cite{martin:2008} is:

\begin{equation}
P_{\mathrm{LH}}~[\mathrm{MW}] = 0.049\, \bar{n}_e^{0.72}\, B_t^{0.8}\, S^{0.94},
\end{equation}

\noindent where $\bar{n}_e$ is the line-averaged electron density in units of $10^{20}~\mathrm{m}^{-3}$, $B_t$ is in T, and $S$ is the plasma surface area in $\mathrm{m}^2$.
The bottom panel of Figure~\ref{fig:plh} shows the peak  powers from the $\dped = 0.03$ and $\dped = 0.05$ scans as a function of $B_t$. From $B_t = 2$--$6$~T, the peak power scales approximately in line with $P_{\mathrm{LH}}$. As $B_t$ increases to 8~T and beyond, the required power begins to deviate from the empirical scaling, and at $B_t = 12$~T the QLGYRO-predicted power to transition from first to second stability rises substantially above it. 

As the plasma shape is the same for these two simulations, very little is actually different in the underlying CGYRO simulations at the different magnetic fields. This is highlighted looking at an example point in table~\ref{tab:cgyro_diff}. By far the biggest difference is simply the gyroBohm normalization flux, being 14~times bigger at 12~T than at 2~T. Recall that $Q_{\mathrm{GB}} = n_e T_e c_s \rho_{*}^2$, which is proportional to $\bpped T_{e,\mathrm{ped}}^{1.5}$ which is proportional to $B_t^{1.5}$ for a given point on the stability map.  This is roughly inline with $P_{\mathrm{LH}} \propto B_t^{1.52}$ in this scan as the density increases with $B_t$ due to holding $f_{\mathrm{GW}}$ fixed.

\begin{table}[h]
\centering
\setlength\tabcolsep{2mm}
\caption{Most different parameters between \texttt{input.cgyro} files in 12~T and 2~T case $\dped=0.035$ and $\bpped=0.09$. Here $L_T = -T_e/(dT_e/dr)$ is the electron temperature gradient scale length.}
\label{tab:cgyro_diff}
\begin{tabular}{lccc}
\hline\hline
Parameter      & 12~T & 2~T  & 12~T/2~T \\
\hline
$Q_{\mathrm{GB}}$       & 0.0013      & 9.2e-05    & 14.1  \\
$\nu_e$         & 0.165     & 0.85     &   0.194 \\
$s$             & 3.5201      & 6.6174      & 0.53 \\
$a/L_{T}$        & 44.346      & 37.779      &  1.17 \\
\hline
\end{tabular}
\end{table}

\begin{figure}
    \centering
    \includegraphics[width=3.5in]{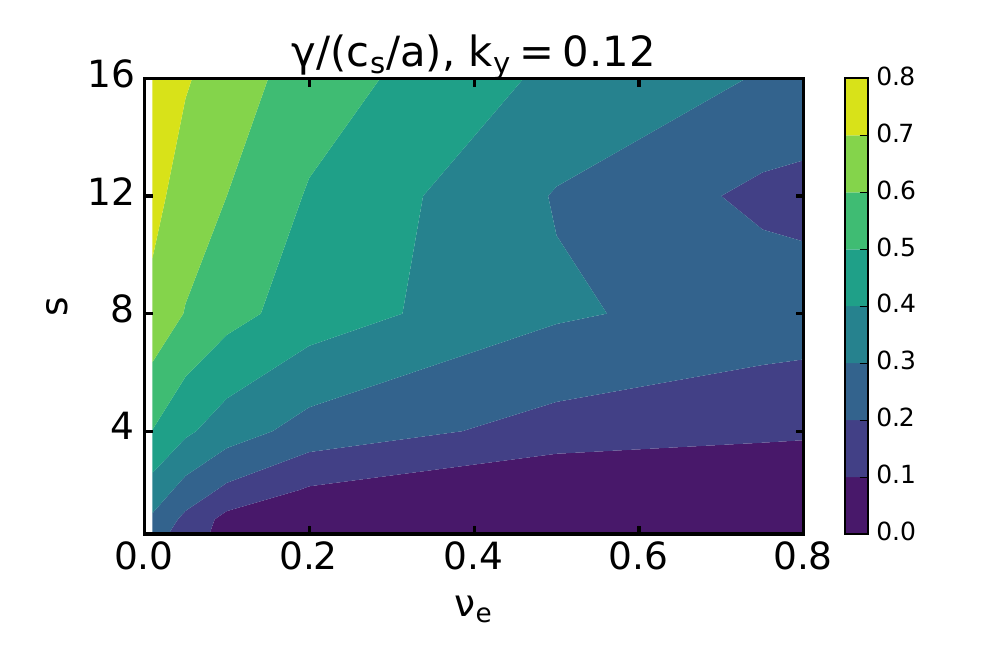}\\
    \caption{CGYRO predicted growth rate of the 2~T $\dped=0.035$ $\bpped=0.09$ scanning $\nu_e$ and $s$.}
    \label{fig:nuscan}
\end{figure}

While the dominant effect of increasing $B_t$ is through the reduction of $Q_{\mathrm{GB}}$, collisionality also significantly modifies the details of the stability map. Collisionality influences the system through two competing mechanisms. First, $\nu_e$ directly affects the KBM growth rates. Second, a reduction in $\nu_e$ increases the bootstrap current efficiency at a given $\bpped$, which in turn reduces the magnetic shear $s$. Figure~\ref{fig:nuscan} shows example CGYRO growth rates obtained by scanning both $\nu_e$ and $s$. The growth rate increases with increasing $s$ and decreases with increasing $\nu_e$, demonstrating that these two effects act in opposition.

\begin{figure}
    \centering
    \includegraphics[width=3.5in]{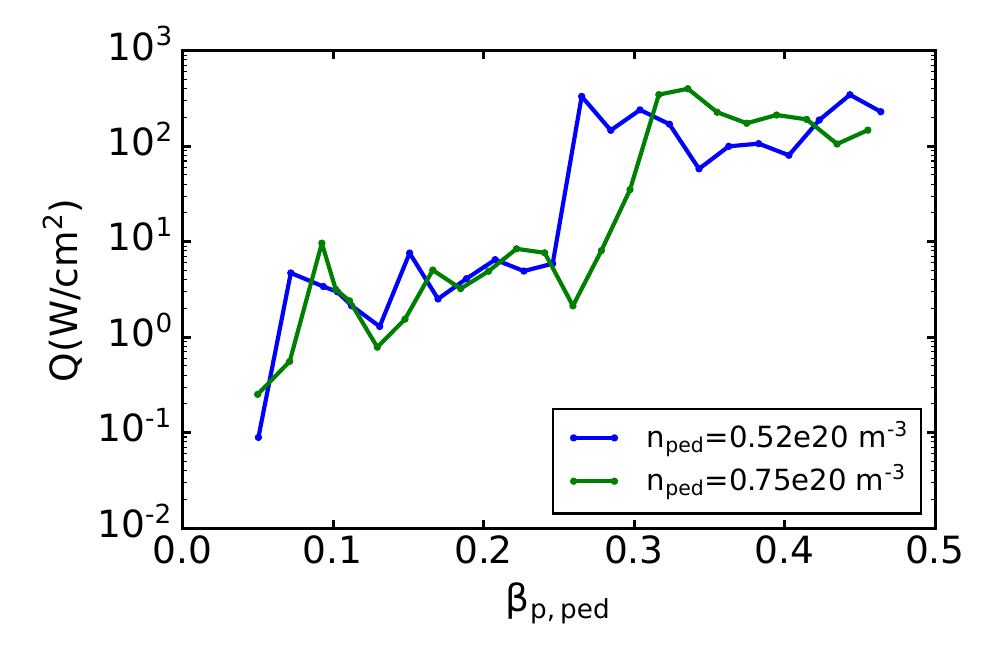}\\
    \caption{QLGYRO predicted heat flux versus $\bpped$ at fixed $\dped = 0.03$ at the nominal density $n_{e,\mathrm{ped}}=0.52 \times 10^{20}~\mathrm{m}^{-3}$ (blue) and increased density of $n_{e,\mathrm{ped}}=0.75 \times 10^{20}~\mathrm{m}^{-3}$ (green).}
    \label{fig:nescan}
\end{figure}

To highlight the importance of collisionality effects on the transport predictions, the QLGYRO $B_t=2$~T simulations are repeated for the $\dped=0.03$ case while increasing only the pedestal density to $n_{e,\mathrm{ped}}=0.75 \times 10^{20}~\mathrm{m}^{-3}$, as shown in Figure~\ref{fig:nescan}. Based on the reduction in $Q_{\mathrm{GB}}$ alone, one would expect a decrease in the heat flux at a given $\bpped$ due to the lower $T_e$, which would be inconsistent with $P_{\mathrm{LH}}$ scalings that generally increase with density. However, the QLGYRO results for this sharp pedestal predict a similar flux level, and even a slightly higher flux at $\bpped=0.1$.  This highlights the importance of including collisionality and magnetic geometry effects in the transport predictions. However, we note that at wider pedestals where the change in $s$ is weaker, $Q_{\mathrm{GB}}$ is more closely followed.

\begin{figure}
    \centering
    \includegraphics[width=3.5in]{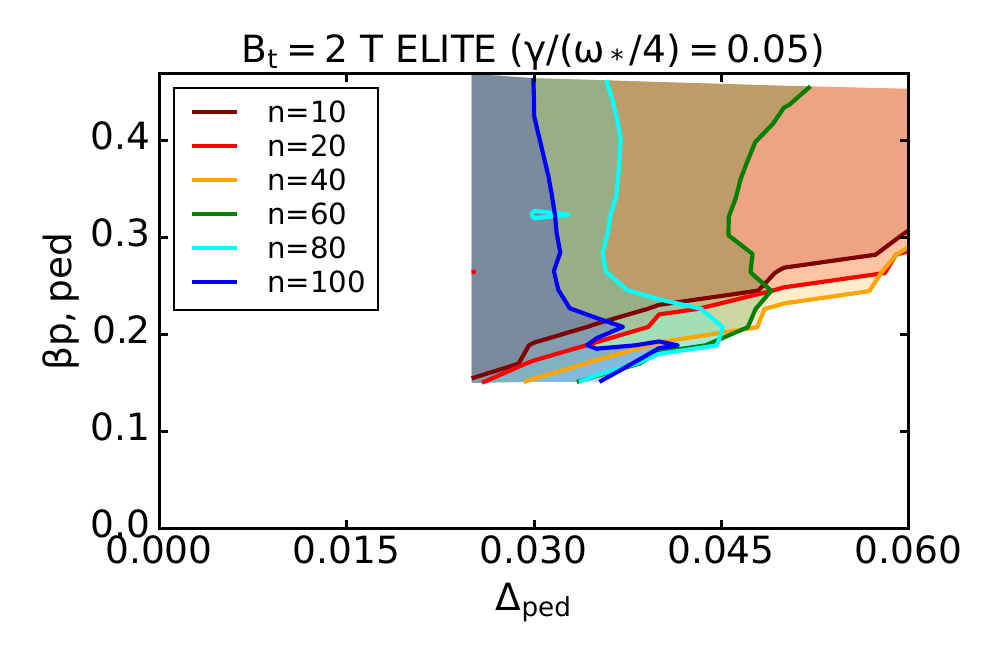} \\
    \caption{ELITE growth rate $\gnorm$ plotted versus $\bpped$ and $\dped$ for $n=10$--$100$ modes.}
    \label{fig:elite_con}
\end{figure}

Lastly, we examine how global effects may influence the $B_t=2$~T case. Figure~\ref{fig:elite_con} shows the ELITE stability thresholds for toroidal mode numbers $n=10$--$100$, using a low instability cutoff of $\gnorm>0.05$ in order to identify the onset of instability. At the steepest pressure gradients, the high- but finite-$n$ ballooning modes are destabilized first. As the pedestal height and width increase further, the high-$n$ modes subsequently restabilize: the $n=100$ mode becomes stable at $\dped=0.04$, the $n=80$ mode at $\dped=0.044$, and the $n=60$ mode at $\dped=0.048$. Interestingly, the onset of second stability in the local KBM calculations occurs in parallel with the global MHD destabilization, as shown in the top panel of Figure~\ref{fig:sparc}. A picture consistent with the EPED model is that KBMs continue to constrain the pedestal height--width scaling, although global effects must also be considered, as suggested by Saarelma~\cite{saarelma:2017}. In this picture, the H-mode pedestal exists in an intermediate regime bounded by the onset of local KBM second stability on one side and global finite-$n$ ballooning instability on the other. However, as the pedestal height and width increase, the high- but finite-$n$ ballooning modes become stabilized, leaving only the low-$n$ peeling modes unstable and ultimately leading to an ELM.

\section{Conclusions}
\label{sec:conclusion}

\noindent In this work, we constructed combined QLGYRO–MHD stability maps by scanning pedestal height and width across a range of aspect ratios, magnetic fields, and triangularities. By integrating gyrokinetic flux predictions with ideal MHD stability boundaries from ELITE and GATO, we obtained a unified picture of pedestal operating space that captures both microinstability-driven transport and macroscopic stability limits.

\noindent Across both spherical and conventional aspect-ratio configurations, a consistent stability structure emerges: an initial first-stability regime at low drive, followed by a KBM-unstable band, a second-stable KBM region, and ultimately limits set by lower-$n$ MHD modes. The H-mode cases examined here appear to lie within the second-stable KBM regime, raising the question of what ultimately constrains the pedestal gradient. One possibility is that modes not included in this analysis, such as ETG turbulence, play a limiting role. Alternatively, our global ELITE finite-$n$ analysis at SPARC parameters shows that finite-$n$ ballooning modes destabilize coincident with the local KBM second-stability onset, suggesting --- consistent with Saarelma~\cite{saarelma:2017} --- that the pedestal is bounded by local KBM second stability below and global finite-$n$ ballooning destabilization above.

\noindent While the absolute flux levels should not be interpreted quantitatively, these results suggest that KBM-mediated transport may provide a useful physics-based framework for understanding relative H-mode accessibility. Scaling the magnetic field while preserving edge safety factor and Greenwald fraction reveals that the qualitative structure of the stability map remains intact, but the QLGYRO predicted KBM flux required to transition from first to second stability increases substantially with $B_t$. The inferred power required to surmount the KBM barrier trends similarly to empirical $P_{\mathrm{LH}}$ scalings primarily due to the increase in the gyroBohm flux normalization. 

\noindent Taken together, these results suggest that pedestal accessibility and performance cannot be understood from a single limiting mechanism. Instead, the interplay between high-$n$ peeling stability, local KBM second stability, global finite-$n$ ballooning instability, and lower-$n$ MHD limits defines a narrow and geometry-dependent operating window. Mapping this coupled stability space provides a practical framework for interpreting present experiments and guiding future pedestal optimization studies.

\begin{acknowledgments}

This material is based upon work supported by the U.S. Department of Energy, Office of Science, Office of Fusion Energy Sciences under Awards DE-SC0021113, DE-SC0024399, and DE-AC02-09CH11466. This research was supported by General Atomics. This work has been part- funded by the EPSRC Energy Programme [grant number EP/W006839/1].  Part of the data analysis was performed using the OMFIT integrated modeling framework \cite{meneghini:2015}. 
\end{acknowledgments}

\spheading{Disclaimer}

This report was prepared as an account of work sponsored by an agency of the United States Government.
Neither the United States Government nor any agency thereof, nor any of their employees, makes any warranty, express or implied, or assumes any legal liability or responsibility for the accuracy, completeness, or usefulness of any information, apparatus, product, or process disclosed, or represents that its use would not infringe privately owned rights. 
Reference herein to any specific commercial product, process, or service by trade name, trademark, manufacturer, or otherwise, does not necessarily constitute or imply its endorsement, recommendation, or favoring by the United States Government or any agency thereof.
The views and opinions of authors expressed herein do not necessarily state or reflect those of the United States Government or any agency thereof.

\bibliographystyle{unsrt}
\bibliography{main.bib}

@STRING{cpc	= "Comput. Phys. Commun." }

@STRING{jcp	= "J. Comput. Phys." }

@STRING{nf	= "Nucl. Fusion" }

@STRING{pop	= "Phys. Plasmas" }

@STRING{ppcf	= "Plasma Phys. Control. Fusion" }

@Article{	  bernard:1981,
  title		= {{GATO: An MHD stability code for axisymmetric plasmas with
		  internal separatrices}},
  author	= {L.C. Bernard and F.J. Helton and R.W. Moore},
  journal	= cpc,
  year		= {1981},
  number	= {3},
  pages		= {377 - 380},
  volume	= {24},
  doi		= {doi.org/10.1016/0010-4655(81)90160-0}
}

@Article{	  candy:2016,
  author	= {J. Candy and E.A. Belli and R.V. Bravenec},
  title		= {A high-accuracy {E}ulerian gyrokinetic solver for
		  collisional plasmas},
  journal	= jcp,
  volume	= 324,
  pages		= 73,
  doi		= {doi.org/10.1016/j.jcp.2016.07.039},
  year		= 2016
}

@Article{	  meneghini:2015,
  author	= {O. Meneghini and S.P. Smith and L.L. Lao and O. Izacard
		  and Q. Ren and J.M. Park and J. Candy and Z. Wang and C.J.
		  Luna and V.A. Izzo and B.A. Grierson and P.B. Snyder and C.
		  Holland and J. Penna and G. Lu and P. Raum and A. McCubbin
		  and D.M. Orlov and E.A. Belli and N.M. Ferraro and R.
		  Prater and T.H. Osborne and A.D. Turnbull and G.M. Staebler
		  and {The AToM Team}},
  title		= {Integrated modeling applications for tokamak experiments
		  with {OMFIT}},
  journal	= nf,
  volume	= 55,
  pages		= 083008,
  year		= 2015,
  doi		= {doi.org/10.1088/0029-5515/55/8/083008}
}

@Article{	  snyder:2002,
  author	= {P. B. Snyder and H. R. Wilson and J.R. Ferron and L. L. Lao
		  and A. W. Leonard and T. H. Osborne and A. D. Turnbull},
  title		= {Edge localized modes and the pedestal: A model based on
		  coupled peeling-ballooning modes},
  journal	= pop,
  volume	= 9,
  pages		= 2037,
  year		= 2002,
  doi		= {doi.org/10.1063/1.1449463}
}

@Article{	  snyder:2011,
  author	= {P.B. Snyder and R.J. Groebner and J.W. Hughes and T.H.
		  Osborne and M. Beurskens and A.W. Leonard and H.R. Wilson
		  and X.Q. Xu},
  title		= {{A first-principles predictive model of the pedestal
		  height and width: development, testing and ITER
		  optimization with the EPED model}},
  journal	= nf,
  volume	= 51,
  pages		= 103016,
  year		= 2011,
  doi		= {doi.org/10.1088/0029-5515/51/10/103016}
}

@Article{	  staebler:2017,
  author	= {G.M. Staebler and N.T. Howard and J. Candy and C.
		  Holland},
  title		= {A model of the saturation of coupled electron and ion
		  scale gyrokinetic turbulence},
  journal	= nf,
  volume	= {57},
  number	= {},
  pages		= {066046},
  year		= {2017},
  doi		= "doi.org/10.1088/1741-4326/aa6bee"
}

@Article{	  wilson:2002,
  author	= {H.R. Wilson and P.B. Snyder and G.T.A. Huysmans},
  title		= {Numerical studies of edge localized instabilities in
		  tokamaks},
  journal	= pop,
  volume	= 9,
  pages		= 1277,
  year		= 2002,
  doi		= {doi.org/10.1063/1.1459058}
}

@article{patel:2021,
year = {2021},
author = {B. Patel},
title = {Confinement physics for a steady state net electric burning spherical tokamak PhD Thesis University of York},
url= {https://etheses.whiterose.ac.uk/28991},
journal={PhD Thesis University of York}
}

@article{parisi:2024,
  author  = {J. F. Parisi and A. O. Nelson and W. Guttenfelder and
             R. Gaur and J. Berkery and S. Kaye and K. Barada and
             C. Clauser and A. Diallo and D. Hatch and A. Kleiner and
             M. Lampert and T. Macwan and J. Menard},
  title   = {Stability and transport of gyrokinetic critical pedestals},
  journal = {Nuclear Fusion},
  volume  = {64},
  pages   = {086034},
  year    = {2024},
  doi     = {10.1088/1741-4326/ad4d02}
}

@article{knolker:2021,
  author  = {M. Knolker and T. Osborne and E. A. Belli and
             S. S. Henderson and A. Kirk and L. Kogan and
             S. Saarelma and P. B. Snyder},
  title   = {Pedestal stability analysis on {MAST} in preparation
             for {MAST-U}},
  journal = {Nuclear Fusion},
  volume  = {61},
  pages   = {046041},
  year    = {2021},
  doi     = {10.1088/1741-4326/abe804}
}

@article{redl:2021,
  author  = {A. Redl and C. Angioni and E. Belli and O. Sauter},
  title   = {A new set of analytical formulas for the current drive
             efficiency and the bootstrap current in toroidal plasmas},
  journal = {Physics of Plasmas},
  volume  = {28},
  pages   = {022502},
  year    = {2021},
  doi     = {10.1063/5.0012664}
}

@article{diallo:2013,
  author    = {Diallo, A. and Kramer, G. J. and Smith, D. R. and Maingi, R. and Bell, R. E. and Guttenfelder, W. and LeBlanc, B. P. and Podest{\`a}, M. and McKee, G. J. and Fonck, R.},
  title     = {Observation of ion scale fluctuations in the pedestal region during the edge-localized-mode cycle on the National Spherical Torus Experiment},
  journal   = pop,
  volume    = {20},
  number    = {1},
  pages     = {012505},
  year      = {2013},
  doi       = {10.1063/1.4773402}
}

@article{slendebroek:2023,
    author = {Slendebroek, T. and McClenaghan, J. and Meneghini, O. M. and Lyons, B. C. and Smith, S. P. and Neiser, T. F. and Shi, N. and Candy, J.},
    title = {Elevating zero dimensional global scaling predictions to self-consistent theory-based simulations},
    journal = pop,
    volume = {30},
    number = {7},
    pages = {072511},
    year = {2023},
    month = {07},
    doi = {10.1063/5.0148886},
}

@article{imada:2024,
doi = {10.1088/1741-4326/ad5219},
year = {2024},
volume = {64},
number = {8},
pages = {086002},
author = {Imada, K. and Osborne, T. H. and Saarelma, S. and Clark, J. G. and Kirk, A. and Knolker, M. and Scannell, R. and Snyder, P.B. and Vincent, C. and Wilson, H.R. and the MAST Upgrade Team},
title = {Observation of a new pedestal stability regime in {MAST} {U}pgrade {H}-mode plasmas},
journal = nf
}

@article{pankin:2025,
doi = {10.1088/1361-6587/ae049c},
url = {https://doi.org/10.1088/1361-6587/ae049c},
year = {2025},
volume = {67},
number = {9},
pages = {095023},
author = {Pankin, Alexei and Ebrahimi, Fatima and King, Jacob and Kleiner, Andreas and Dominguez-Palacios, Jesus},
title = {Effects beyond ideal {MHD} on stability of wide and enhanced pedestal regimes in NSTX},
journal = ppcf,
}

@article{liu:2025,
    author = {Liu, Yueqiang and Zhao, Chen and Ebrahimi, Fatima},
    title = {Peeling-ballooning modes in spherical tokamaks: {M}ulti-branch instabilities and effects beyond ideal {MHD}},
    journal = pop,
    volume = {32},
    number = {12},
    pages = {122507},
    year = {2025},
    doi = {10.1063/5.0305045},
}

@article{kleiner:2025,
doi = {10.1088/1361-6587/adf91a},
year = {2025},
volume = {67},
number = {8},
pages = {085026},
author = {Kleiner, A and Imada, K and Ebrahimi, F and Ferraro, N M and Haskey, S R and Kogan, L and Pankin, A},
title = {A study of resistive peeling–ballooning modes across spherical tokamaks},
journal = ppcf,
}

@article{osborne:2015,
doi = {10.1088/0029-5515/55/6/063018},
year = {2015},
volume = {55},
number = {6},
pages = {063018},
author = {Osborne, T.H. and Jackson, G.L. and Yan, Z. and Maingi, R. and Mansfield, D.K. and Grierson, B.A. and Chrobak, C.P. and McLean, A.G. and Allen, S.L. and Battaglia, D.J. and Briesemeister, A.R. and Fenstermacher, M.E. and McKee, G.R. and Snyder, P.B. and The DIII-D Team},
title = {Enhanced {H}-mode pedestals with lithium injection in {DIII-D}},
journal = nf,
}

@article{harrison:2019,
doi = {10.1088/1741-4326/ab121c},
year = {2019},
volume = {59},
number = {11},
pages = {112011},
author = {Harrison, J.R. and Akers, R.J. and Allan, S.Y. and Allcock, J.S. and Allen, J.O. and Appel, L. and Barnes, M. and Ben Ayed, N. and Boeglin, W. and Bowman, C. and Bradley, J. and Browning, P. and Bryant, P. and Carr, M. and Cecconello, M. and Challis, C.D. and Chapman, S. and Chapman, I.T. and Colyer, G.J. and Conroy, S. and Conway, N.J. and Cox, M. and Cunningham, G. and Dendy, R.O. and Dorland, W. and Dudson, B.D. and Easy, L. and Elmore, S.D. and Farley, T. and Feng, X. and Field, A.R. and Fil, A. and Fishpool, G.M. and Fitzgerald, M. and Flesch, K. and Fox, M.F.J. and Frerichs, H. and Gadgil, S. and Gahle, D. and Garzotti, L. and Ghim, Y.-C. and Gibson, S. and Gibson, K.J. and Hall, S. and Ham, C. and Heiberg, N. and Henderson, S.S. and Highcock, E. and Hnat, B. and Howard, J. and Huang, J. and Irvine, S.W.A. and Jacobsen, A.S. and Jones, O. and Katramados, I. and Keeling, D. and Kirk, A. and Klimek, I. and Kogan, L. and Leland, J. and Lipschultz, B. and Lloyd, B. and Lovell, J. and Madsen, B. and Marshall, O. and Martin, R. and McArdle, G. and McClements, K. and McMillan, B. and Meakins, A. and Meyer, H.F. and Militello, F. and Milnes, J. and Mordijck, S. and Morris, A.W. and Moulton, D. and Muir, D. and Mukhi, K. and Murphy-Sugrue, S. and Myatra, O. and Naylor, G. and Naylor, P. and Newton, S.L. and O’Gorman, T. and Omotani, J. and O’Mullane, M.G. and Orchard, S. and Pamela, S.J.P. and Pangione, L. and Parra, F. and Perez, R.V. and Piron, L. and Price, M. and Reinke, M.L. and Riva, F. and Roach, C.M. and Robb, D. and Ryan, D. and Saarelma, S. and Salewski, M. and Scannell, S. and Schekochihin, A.A. and Schmitz, O. and Sharapov, S. and Sharples, R. and Silburn, S.A. and Smith, S.F. and Sperduti, A. and Stephen, R. and Thomas-Davies, N.T. and Thornton, A.J. and Turnyanskiy, M. and Valovič, M. and Van Wyk, F. and Vann, R.G.L. and Walkden, N.R. and Waters, I. and Wilson, H.R. and the MAST-U Team and the EUROfusion MST1 Team},
title = {Overview of new {MAST} physics in anticipation of first results from {MAST} {U}pgrade},
journal = nf

}

@article{ono:2000,
doi = {10.1088/0029-5515/40/3Y/316},
year = {2000},
volume = {40},
number = {3Y},
pages = {557},
author = {M. Ono and S.M. Kaye and Y.-K.M. Peng and G. Barnes and W. Blanchard and M.D. Carter and J. Chrzanowski and L. Dudek and R. Ewig and D. Gates and R.E. Hatcher and T. Jarboe and S.C. Jardin and D. Johnson and R. Kaita and M. Kalish and C.E. Kessel and H.W. Kugel and R. Maingi and R. Majeski and J. Manickam and B. McCormack and J. Menard and D. Mueller and B.A. Nelson and B.E. Nelson and C. Neumeyer and G. Oliaro and F. Paoletti and R. Parsells and E. Perry and N. Pomphrey and S. Ramakrishnan and R. Raman and G. Rewoldt and J. Robinson and A.L. Roquemore and P. Ryan and S. Sabbagh and D. Swain and E.J. Synakowski and M. Viola and M. Williams and J.R. Wilson and NSTX Team},
title = {Exploration of spherical torus physics in the {NSTX} device},
journal = nf
}

@article{battaglia:2020,
    author = {Battaglia, D. J. and Guttenfelder, W. and Bell, R. E. and Diallo, A. and Ferraro, N. and Fredrickson, E. and Gerhardt, S. P. and Kaye, S. M. and Maingi, R. and Smith, D. R.},
    title = {Enhanced pedestal {H}-mode at low edge ion collisionality on {NSTX}},
    journal = pop,
    volume = {27},
    number = {7},
    pages = {072511},
    year = {2020},
    month = {07},
    doi = {10.1063/5.0011614},
}

@article{luxon:2002,
doi = {10.1088/0029-5515/42/5/313},
year = {2002},
volume = {42},
number = {5},
pages = {614},
author = {J.L. Luxon},
title = {A design retrospective of the {DIII-D} tokamak},
journal = nf
}

@article{keilhacker:1985,
doi = {10.1088/0029-5515/25/9/008},
year = {1985},
volume = {25},
number = {9},
pages = {1045},
author = {Keilhacker, M. and ASDEX Team},
title = {The {ASDEX} divertor tokamak},
journal = nf
}

@article{martin:2008,
doi = {10.1088/1742-6596/123/1/012033},
year = {2008},
volume = {123},
number = {1},
pages = {012033},
author = {Y R Martin and T Takizuka and (and the ITPA CDBM H-mode Threshold Database Working Group)},
title = {Power requirement for accessing the H-mode in ITER},
journal = {Journal of Physics: Conference Series},
}

@article{saarelma:2017,
doi = {10.1088/1361-6587/aa66ab},
year = {2017},
volume = {59},
number = {6},
pages = {064001},
author = {Saarelma, S and Martin-Collar, J and Dickinson, D and McMillan, B F and Roach, C M and MAST team and The JET Contributors4},
title = {Non-local effects on pedestal kinetic ballooning mode stability},
journal = ppcf,

}

@article{miller:1998,
    author = {Miller, R. L. and Chu, M. S. and Greene, J. M. and Lin-Liu, Y. R. and Waltz, R. E.},
    title = {Noncircular, finite aspect ratio, local equilibrium model},
    journal = pop,
    volume = {5},
    number = {4},
    pages = {973-978},
    year = {1998},
    doi = {10.1063/1.872666},
}

@article{rodriguez:2022,
doi = {10.1088/1741-4326/ac1654},
year = {2022},
volume = {62},
number = {4},
pages = {042003},
author = {Rodriguez-Fernandez, P. and Creely, A. J. and Greenwald, M. J. and Brunner, D. and Ballinger, S. B. and Chrobak, C. P. and Garnier, D. T. and Granetz, R. and Hartwig, Z. S. and Howard, N. T. and Hughes, J. W. and Irby, J. H. and Izzo, V. A. and Kuang, A. Q. and Lin, Y. and Marmar, E. S. and Mumgaard, R. T. and Rea, C. and Reinke, M. L. and Riccardo, V. and Rice, J. E. and Scott, S. D. and Sorbom, B. N. and Stillerman, J. A. and Sweeney, R. and Tinguely, R. A. and Whyte, D. G. and Wright, J. C. and Yuryev, D. V.},
title = {Overview of the {SPARC} physics basis towards the exploration of burning-plasma regimes in high-field, compact tokamaks},
journal = nf,
}

\end{document}